\documentclass[aps,prl, reprint, floatfix, superscriptaddress, showpacs]{revtex4-1}

\usepackage{graphicx}
\begin{document}

\preprint{}

\title{Reaffirming the $d_{x^2-y^2}$ Superconducting Gap Using the Autocorrelation Angle-Resolved Photoemission Spectroscopy of Bi$_{1.5}$Pb$_{0.55}$Sr$_{1.6}$La$_{0.4}$CuO$_{6+\delta}$}

\author{M. Hashimoto}
\affiliation{Stanford Synchrotron Radiation Lightsource, SLAC National Accelerator Laboratory, 2575, Sand Hill Road, Menlo Park, California 94025, USA}
\affiliation{Stanford Institute for Materials and Energy Sciences, SLAC National Accelerator Laboratory, 2575 Sand Hill Road, Menlo Park, California 94025, USA}
\affiliation{Geballe Laboratory for Advanced Materials, Departments of Physics and Applied Physics, Stanford University, Stanford, California 94305, USA}
\affiliation{Advanced Light Source, Lawrence Berkeley National Lab, Berkeley, California 94720, USA}
\author{R.-H. He}
\affiliation{Stanford Institute for Materials and Energy Sciences, SLAC National Accelerator Laboratory, 2575 Sand Hill Road, Menlo Park, California 94025, USA}
\affiliation{Geballe Laboratory for Advanced Materials, Departments of Physics and Applied Physics, Stanford University, Stanford, California 94305, USA}
\affiliation{Advanced Light Source, Lawrence Berkeley National Lab, Berkeley, California 94720, USA}
\author{J. P. Testaud}
\affiliation{Stanford Institute for Materials and Energy Sciences, SLAC National Accelerator Laboratory, 2575 Sand Hill Road, Menlo Park, California 94025, USA}
\affiliation{Geballe Laboratory for Advanced Materials, Departments of Physics and Applied Physics, Stanford University, Stanford, California 94305, USA}
\affiliation{Advanced Light Source, Lawrence Berkeley National Lab, Berkeley, California 94720, USA}
\author{W. Meevasana}
\affiliation{Stanford Institute for Materials and Energy Sciences, SLAC National Accelerator Laboratory, 2575 Sand Hill Road, Menlo Park, California 94025, USA}
\affiliation{Geballe Laboratory for Advanced Materials, Departments of Physics and Applied Physics, Stanford University, Stanford, California 94305, USA}
\affiliation{School of Physics, Suranaree University of Technology, Nakhon Ratchasima, 30000 Thailand}
\author{R. G. Moore}
\affiliation{Stanford Synchrotron Radiation Lightsource, SLAC National Accelerator Laboratory, 2575, Sand Hill Road, Menlo Park, California 94025, USA}
\affiliation{Stanford Institute for Materials and Energy Sciences, SLAC National Accelerator Laboratory, 2575 Sand Hill Road, Menlo Park, California 94025, USA}
\affiliation{Geballe Laboratory for Advanced Materials, Departments of Physics and Applied Physics, Stanford University, Stanford, California 94305, USA}
\author{D. H. Lu}
\affiliation{Stanford Synchrotron Radiation Lightsource, SLAC National Accelerator Laboratory, 2575, Sand Hill Road, Menlo Park, California 94025, USA}
\author{Y. Yoshida}
\affiliation{Nanoelectronics Research Institute, AIST, Ibaraki 305-8568, Japan}
\author{H. Eisaki}
\affiliation{Nanoelectronics Research Institute, AIST, Ibaraki 305-8568, Japan}
\author{T. P. Devereaux}
\affiliation{Stanford Institute for Materials and Energy Sciences, SLAC National Accelerator Laboratory, 2575 Sand Hill Road, Menlo Park, California 94025, USA}
\affiliation{Geballe Laboratory for Advanced Materials, Departments of Physics and Applied Physics, Stanford University, Stanford, California 94305, USA}
\author{Z. Hussain}
\affiliation{Advanced Light Source, Lawrence Berkeley National Lab, Berkeley, California 94720, USA}
\author{Z.-X. Shen}
\affiliation{Stanford Institute for Materials and Energy Sciences, SLAC National Accelerator Laboratory, 2575 Sand Hill Road, Menlo Park, California 94025, USA}
\affiliation{Geballe Laboratory for Advanced Materials, Departments of Physics and Applied Physics, Stanford University, Stanford, California 94305, USA}

\date{\today}

\begin{abstract}
Knowledge of the gap function is important to understand the pairing mechanism for high-temperature ($T_\mathrm{c}$) superconductivity. However, Fourier transform scanning tunneling spectroscopy (FT STS) and angle-resolved photoemission spectroscopy (ARPES) in the cuprates have reported contradictory gap functions, with FT-STS results strongly deviating from a canonical $d_{x^2-y^2}$ form. By applying an ``octet model'' analysis to autocorrelation ARPES, we reveal that a contradiction occurs because the octet model does not consider the effects of matrix elements and the pseudogap. This reaffirms the canonical $d_{x^2-y^2}$ superconducting gap around the node, which can be directly determined from ARPES. Further, our study suggests that the FT-STS reported fluctuating superconductivity around the node at far above $T_\mathrm{c}$ is not necessary to explain the existence of the quasiparticle interference.
\end{abstract}


\maketitle

The superconducting gap magnitude in a superconductor reflects the pairing strength of a Cooper pair, which is closely related to the transition temperature ($T_\mathrm{c}$). Because of that, the experimental determination of the gap magnitude is vitally important to understand the mechanism of superconductivity. In a high-$T_\mathrm{c}$ cuprate superconductor, the momentum ($\mathbf{k}$) dependence of the gap has to be determined at the same time. However, two complementary leading tools for such information, angle-resolved photoemission spectroscopy (ARPES) and Fourier transform scanning tunneling spectroscopy (FT STS), have reported inconsistent gap functions. While all ARPES studies have consistently indicated the presence of nodal $d_{x^2-y^2}$ excitations below $T_\mathrm{c}$ \cite{TanakaDistinct2006, LeeAbrupt2007, KondoCompetition2009, KanigelProtected2007, ChatterjeeObservation2010, DamascelliAngle-resolved2003}, FT-STS studies have reported finite gapless ``Fermi arcs'' \cite{HoffmanImaging2002, McElroyRelating2003, HanaguriQuasiparticle2007, WiseImaging2009, KohsakaHow2008}, which persists well above $T_\mathrm{c}$ with no signature across the superconducting transition \cite{LeeSpectroscopic2009}. These momentum and temperature dependences of a FT-STS derived gap function also conflict with transport, and suggest a more complex nature for high-$T_\mathrm{c}$ superconductivity. Given the high influence of these experimental tools and the issue of whether the superconducting fluctuation extends well above $T_\mathrm{c}$, it is important to ascertain the gap functional form for a successful development of a microscopic theory.

STS is a real-space ($\mathbf{r}$-space) probe that can visualize $\mathbf{k}$-space properties via FT STS. When quasiparticles scatter off impurities, a quasiparticle standing wave interference (QPI) pattern is formed in $\mathbf{r}$-space, that modulates the density of states and is most clearly displayed in momentum transfer ($\mathbf{q}$) space via the FT of the $\mathbf{r}$-space image. In the octet model for $d$-wave superconductors \cite{WangQuasiparticle2003, MarkiewiczBridging2004, WulinModel2009, WulinContrasting}, eight $\mathbf{k}$-space points at the same energy on the Fermi surface (FS) within the first Brillouin zone are connected by seven vectors in momentum transfer ($\mathbf{q}$) space $\mathbf{q}_1$ to $\mathbf{q}_7$, which are associated with the superconducting gap size at the respective $\mathbf{k}$-space points \cite{HoffmanImaging2002, McElroyRelating2003, HanaguriQuasiparticle2007, WiseImaging2009, KohsakaHow2008, LeeSpectroscopic2009}. We emphasize that FT-STS is reliant on the octet model to obtain the gap function, in contrast to the direct determination of the gap by ARPES.

It has been reported that STS matrix elements suppress the nodal region in $\mathbf{k}$ space \cite{AndersenLDA1995, WuAbsence2000,MartinImpurity2002, NieminenOrigin2009}. This could explain why the FT-STS gap function obtained from octet-model analysis lacks information around the node \cite{HoffmanImaging2002, McElroyRelating2003, HanaguriQuasiparticle2007, WiseImaging2009, KohsakaHow2008, LeeSpectroscopic2009}. However, such matrix element effects have not been studied extensively because they are difficult to determine experimentally. A possible way to approach this issue is to investigate the autocorrelation (AC) of ARPES intensities in $\mathbf{q}$ space $I_{AC}(\mathbf{q} ,\omega) = \int I(\mathbf{k}, \omega)I(\mathbf{k}+\mathbf{q}, \omega)d\mathbf{k}$, which has been interpreted as joint density of states (JDOS) $\int A(\mathbf{k}, \omega)A(\mathbf{k}+\mathbf{q}, \omega)d\mathbf{k}$. Here, ARPES intensity is $I(\mathbf{k}, \omega) = M(\mathbf{k}, h\nu, \mathbf{A})A(\mathbf{k}, \omega)f(\omega , T)$ where $M(\mathbf{k}, h\nu, \mathbf{A})$ are matrix elements, $A(\mathbf{k}, \omega)$ is the spectral function, and $f(\omega , T)$ is the Fermi-Dirac function. While not the same, the theoretical similarity between JDOS and QPI has been pointed out \cite{WangQuasiparticle2003, MarkiewiczBridging2004, WulinModel2009, WulinContrasting} and the experimental comparisons between FT STS and AC ARPES have supported it \cite{McElroyElastic2006, ChatterjeeNondispersive2006}.

In this Letter, to obtain AC-ARPES intensities, we use an ARPES data set of nearly optimally doped Bi$_{1.5}$Pb$_{0.55}$Sr$_{1.6}$La$_{0.4}$CuO$_{6+\delta}$ (Pb-Bi2201, $T_\mathrm{c}$ = 38 K, the pseudogap temperature $T^*$ $\sim $ 140 K) \cite{SupACARPES}. Here we stress that we choose an experimental condition where the nodal spectral intensities are suppressed by matrix elements \cite{SupACARPES}, utilizing the fact that matrix elements highly depend on incident photon energy and polarization \cite{BansilImportance99}. This is similar to the STS theoretical expectation \cite{AndersenLDA1995, WuAbsence2000,MartinImpurity2002, NieminenOrigin2009} and thus allows us to discuss the STS observations, especially the FT-STS derived gap function. We show that the obtained AC-ARPES intensities are severely affected by the convoluted effect of intrinsic spectral widths, the pseudogap, and matrix elements. Eventually, applying the octet model, which does not take such effects into consideration, induces a form of the gap function that differs from ARPES but is very similar to the FT-STS \cite{HoffmanImaging2002, McElroyRelating2003, HanaguriQuasiparticle2007, WiseImaging2009, KohsakaHow2008} and temperature-dependent FT-STS \cite{LeeSpectroscopic2009} results. Our result reaffirms the $d_{x^2-y^2}$ superconducting gap function near the node below $T_\mathrm{c}$, which is consistent with previous ARPES studies \cite{TanakaDistinct2006, LeeAbrupt2007, KondoCompetition2009, KanigelProtected2007, ChatterjeeObservation2010, DamascelliAngle-resolved2003}.

\begin{figure}
\begin{center}
\includegraphics[width=0.8\linewidth]{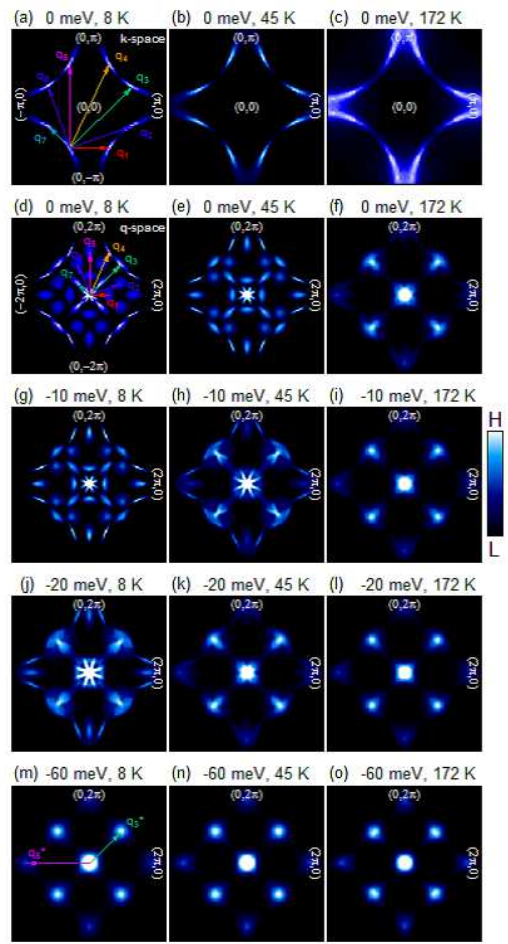}
\caption{(Color online) Image plots in $\mathbf{k}$- and $\mathbf{q}$-spaces of nearly optimally-doped Pb-Bi2201. (a)-(c) $\mathbf{k}$-space ARPES intensities at 0 meV ($E_\mathrm{F}$) for three temperatures, 8 K ($<T_\mathrm{c}$), 45 K ($T_\mathrm{c} < T \ll T^*$) and 172 K ($>T^*$), respectively. (d)-(o), $\mathbf{q}$-space AC-ARPES intensities at various energies for the three temperatures as denoted. All the intensities are integrated within $\pm $0.5 meV. $\mathbf{q}_1-\mathbf{q}_7$ in $\mathbf{q}$-space are exemplified in (d) in appropriate colors, which are the $\mathbf{q}$-vectors expected for the octet model. Corresponding $\mathbf{q}$ vectors in $\mathbf{k}$ space are indicated in (a). Definitions of $\mathbf{q}_3^*$ and $\mathbf{q}_5^*$ at higher binding energy in $\mathbf{q}$ space are shown in (m).
}
\label{Fig1}
\end{center}
\end{figure}

The $\mathbf{k}$-space ARPES intensities of Pb-Bi2201 at $E_\mathrm{F}$ at different temperatures are shown in Figs. \ref{Fig1}(a)-(c). At 172 K ($>T^*$), where no gap exists, the intensity near the node $\sim$($\pi $/2, $\pi $/2) is very weak compared to the antinodal region $\sim$($\pi $, 0) intensity. This confirms that the present experimental condition suppresses the nodal region intensity similar to the STS theoretical expectation \cite{AndersenLDA1995, WuAbsence2000,MartinImpurity2002, NieminenOrigin2009}. From these $\mathbf{k}$-space ARPES intensities with strong matrix element effects, we obtain the $\mathbf{q}$-space AC-ARPES intensities at different energies as shown in Figs. \ref{Fig1}(d)-(o). At 8 K ($\ll T_\mathrm{c}$), there exist high-intensity spots which can be attributed to the octet-model $\mathbf{q}$ vectors $\mathbf{q}_1$--$\mathbf{q}_7$ at low binding energies ($\lesssim -$20 meV). Here we define $\mathbf{q}_1$--$\mathbf{q}_7$ as indicated in Fig. \ref{Fig1}(d), and these vectors can be associated with the vectors between the high-intensity spots in $\mathbf{k}$ space as exemplified in Fig. \ref{Fig1}(a). At higher binding energy, the existence of $\mathbf{q}_1$--$\mathbf{q}_7$ becomes less clear and, instead, the broader spots defined as $\mathbf{q}_3^*$ and $\mathbf{q}_5^*$ appear[Fig. \ref{Fig1}(m)]. Upon raising the temperature above $T_\mathrm{c}$, one can still see $\mathbf{q}_1$--$\mathbf{q}_7$ as shown in Fig. \ref{Fig1}(e), but they completely disappear above $T^*$. On the other hand, $\mathbf{q}_3^*$ and $\mathbf{q}_5^*$ show little temperature dependence. At $T > T^*$ where neither the superconductivity nor pseudogap exists, one can only find $\mathbf{q}_3^*$ and $\mathbf{q}_5^*$ at all the energies [Fig. \ref{Fig1}(f), (i), (l) and (o)]. 

\begin{figure}
\begin{center}
\includegraphics[width=0.75\linewidth]{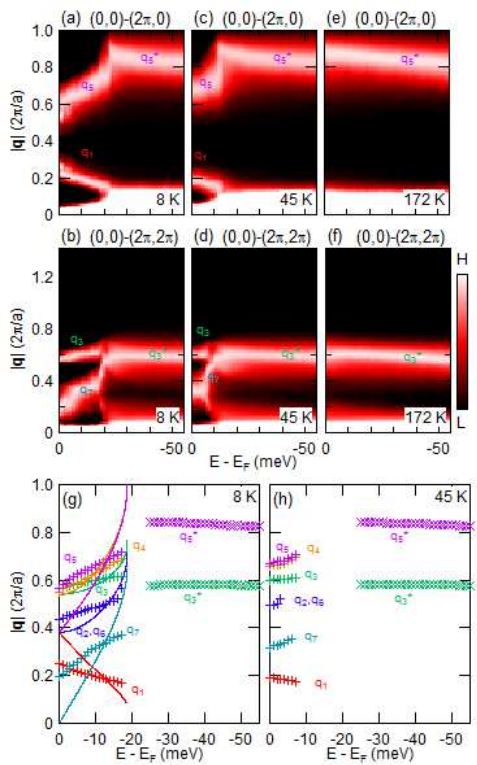}
\caption{(Color online) Energy dispersions of the $\mathbf{q}$-vectors from AC-ARPES. (a)-(f) Energy dependences of AC-ARPES intensities along high-symmetry $\mathbf{q}$ directions at 8 K ($<T_\mathrm{c}$), 45 K ($T_\mathrm{c}<T\ll T^*$), and 172 K ($>T^*$), as denoted. The intensities at each energy are normalized to the height of $\mathbf{q}_5$ (or $\mathbf{q}_5^*$) in (a), (c) and (e), and $\mathbf{q}_3$ (or $\mathbf{q}_3^*$) in (b), (d), and (f), respectively, for visualization purposes. (g),(h) Summary of the extracted dispersions of the $\mathbf{q}$ vectors at 8 K ($<T_\mathrm{c}$) and 45 K ($T_\mathrm{c}<T\ll T^*$), respectively. Solid curves in (g) are the dispersions of $\mathbf{q}_1$--$\mathbf{q}_7$ for a $d_{x^2-y^2}$ gap $\Delta_{\mathrm{sc}}(\mathbf{k})= \Delta _{\mathrm{max}}|\cos(k_x)-\cos(k_y)|/2$ ($\Delta _{\mathrm{max}}$ = 19 meV). The same colors as Fig. \ref{Fig1} are used for the symbols and labels of the $\mathbf{q}$ vectors. 
}
\label{Fig2}
\end{center}
\end{figure}

Next, we extract the energy dispersion of the $\mathbf{q}$-vectors \cite{SupACARPES}. In the AC-ARPES high-symmetry cuts below $T_\mathrm{c}$ shown in Figs. \ref{Fig2}(a) and (b), $\mathbf{q}_1$--$\mathbf{q}_7$ show clear dispersion ($\gtrsim -$ 18 meV) as expected in the octet model. At higher binding energy ($<-$30 meV), $\mathbf{q}_3^*$ and $\mathbf{q}_5^*$ replace $\mathbf{q}_1$--$\mathbf{q}_7$. $\mathbf{q}_1$--$\mathbf{q}_7$ show dispersion also at 45 K ($>T_\mathrm{c}$) with small change of $|\mathbf{q}|$ from those at 10 K, but become harder to track at higher binding energy ($< -$8 meV) as shown in Figs. \ref{Fig2}(c) and (d). Notably, Figs. \ref{Fig2}(e) and (f) reveal that, even above $T^*$, $\mathbf{q}_3^*$ and $\mathbf{q}_5^*$ dominate for all energies. These $\mathbf{q}$ vectors show weak but finite dispersions reflecting the bare band structure. Although the overall features look similar, this is different from the FT-STS observation of the energy independent $\mathbf{q}$ vectors at higher energy below $T^*$. They have been associated with the density-wave-like pseudogap \cite{Hoffmanfour2002, WiseCharge-density-wave2008}.

\begin{figure}
\begin{center}
\includegraphics[width=0.85\linewidth]{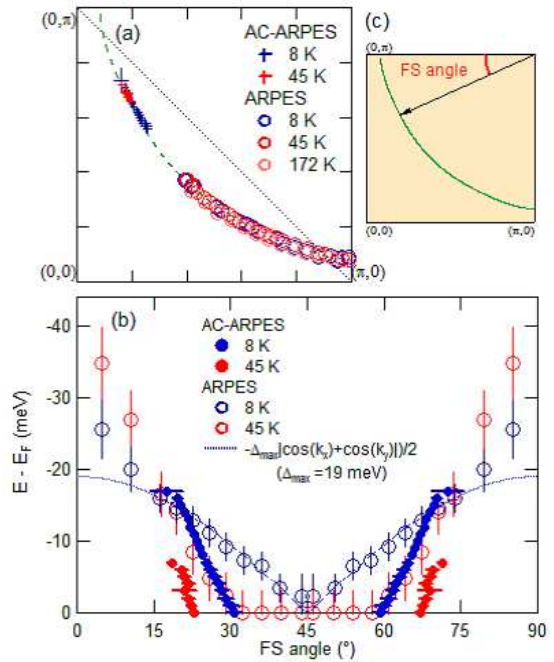}
\caption{(Color online) Deviation from the $d_{x^2-y^2}$ gap anisotropy in the octet-model analysis of AC ARPES. (a) Comparison of the FS shape obtained from the two different analysis methods applied to the same data at various temperatures. Dashed curve and dotted line are guides for the eye for the FS shape and the antiferromagnetic Brillouin zone boundary, respectively. (b) Comparison of the gap anisotropy between the two methods. The definition of the FS angle is shown in (c). Dotted curve in (b) is for the $d_{x^2-y^2}$ gap function ($\Delta _{\mathrm{max}}$ = 19 meV). The energy ranges for the FS shape from AC ARPES in (a) are the same as those for the gap functions in (b).
}
\label{Fig3}
\end{center}
\end{figure}

Applying the octet-model analysis to the extracted $\mathbf{q}_1$--$\mathbf{q}_7$ at 10 and 45 K, as summarized in Fig. \ref{Fig2}(g) and (h), the gap function and the minimum gap loci (MGLs) which reflect the FS shape can be reconstructed and compared to the results obtained from a direct ARPES analysis \cite{SupACARPES}. As shown in Fig. \ref{Fig3}(a), the two methods show a consistent FS shape for all temperatures although the MGLs from AC ARPES are limited to a more narrow region of $\mathbf{k}$ space. This suggests that $\mathbf{q}_1$--$\mathbf{q}_7$ track the electronic properties approximately along the FS and the observation of $\mathbf{q}_1$--$\mathbf{q}_7$ is because of the existence of a gap about $E_\mathrm{F}$ as they are not observed above $T^*$ [Figs. \ref{Fig2}(e) and (f)]. However, the gap anisotropies from the two methods show apparently different behaviors as seen in Fig. \ref{Fig3}(b). At $T < T_\mathrm{c}$, the ARPES derived gap anisotropies show a simple canonical $d_{x^2-y^2}$ form with point nodes at the zone diagonal consistent with the previous ARPES studies \cite{TanakaDistinct2006, LeeAbrupt2007, KondoCompetition2009, KanigelProtected2007, ChatterjeeObservation2010, DamascelliAngle-resolved2003}, while the octet-model analysis suggests the absence of the gap near the node (formation of a Fermi arc). In addition, the gap function near the antinode is lost. The AC-ARPES derived gap anisotropy does not abruptly  change when crossing $T_\mathrm{c}$ from below. The ungapped Fermi arc region becomes larger but the overall feature is maintained. This is well contrasted with the abrupt gap closing and the formation of the Fermi arc above $T_\mathrm{c}$ in the ARPES derived gap function. While the two methods show inconsistent gap functions, the AC-ARPES derived gap function is very similar to the one from FT STS \cite{HoffmanImaging2002, McElroyRelating2003, HanaguriQuasiparticle2007, WiseImaging2009, KohsakaHow2008, LeeSpectroscopic2009} both in terms of a momentum and temperature dependences \cite{LeeSpectroscopic2009}.

We first discuss the most critical difference of the gap function around the node between the two methods. In the octet model, the canonical $d_{x^2-y^2}$ gap function expects the merger of $\mathbf{q}_1$--$\mathbf{q}_7$ into 3 vectors at $E_\mathrm{F}$ as indicated by the solid curves in Fig. \ref{Fig2}(g), as the gap is closed only at the node. This means that there exist at $E_\mathrm{F}$ only 4 high-intensity spots at the nodes in the $\mathbf{k}$-space image plot. However, even though we observed a gap closing only at the node from ARPES \cite{SupACARPES}, simultaneously, we clearly observed 8 high-intensity spots away from the node at $E_\mathrm{F}$ as shown in Fig. \ref{Fig1}(a). They generate the 7 $\mathbf{q}$ vectors in the AC-ARPES intensities [Fig. \ref{Fig1}(d)], and consequently, the MGL and gap function terminate before reaching the node, giving an apparent Fermi arc-like feature even below $T_\mathrm{c}$.

These seemingly contradictory observations in the same data (the gap closure at the node and the 8 high-intensity spots at $E_\mathrm{F}$ in $\mathbf{k}$ space) can be understood by considering the compounded effect of  intrinsic spectral widths and matrix elements. Because of the spectral widths by lifetime broadening due to intrinsic and extrinsic scattering, outside the node, ARPES can have finite intensities at $E_\mathrm{F}$. If the $\mathbf{k}$ dependence of the matrix elements is strong enough as in the present experimental condition or STS theoretical expectation \cite{AndersenLDA1995, WuAbsence2000,MartinImpurity2002, NieminenOrigin2009}, these intensities away from the node can be stronger than those at the nodes, and eventually give 8 high-intensity spots even at $E_\mathrm{F}$.

To conclude, as we demonstrated, the $\mathbf{q}$-space spectra both from AC ARPES and FT STS give an incorrect gap function, because the octet model relies on the $\mathbf{q}$-space intensity variation and ignores the effects of intrinsic spectral widths and matrix elements. The similar but inaccurate gap function from AC ARPES and FT STS may be understood as a result of similar matrix elements. By manipulating matrix elements, it is further confirmed that the octet-model gap function is strongly influenced by matrix elements \cite{SupACARPES}. On the other hand, the gap function directly determined from ARPES does not suffer from matrix elements because each ARPES spectrum at a fixed momentum is evaluated individually and the relative intensities between different momenta do not affect the gap determination.

Our results suggest that the gap function in the antinodal region is affected by the existence of the pseudogap in addition to the effects discussed above. Interestingly, $\mathbf{q}_1$--$\mathbf{q}_7$ are clearly observed even slightly above $T_\mathrm{c}$, similar to the recent FT-STS report \cite{LeeSpectroscopic2009}. This suggests that $\mathbf{q}_1$--$\mathbf{q}_7$ are affected by the pseudogap physics, indicating that the observation of $\mathbf{q}_1$--$\mathbf{q}_7$ is not sufficient to conclude whether the pseudogap is fluctuating superconductivity or competing order. Further, as shown in Fig. \ref{Fig3}(b), the AC-ARPES gap function changes continuously across $T_\mathrm{c}$, suggesting that the gap function from AC ARPES in the superconducting state [Fig. \ref{Fig3}(b)] is also affected by the existence of the pseudogap, but the effect of the superconducting gap that opens on the Fermi arc is not strong. We may conclude that the existence of the QPI above $T_\mathrm{c}$ does not necessarily require fluctuating superconductivity near the node, but could be consistent with fluctuating superconductivity in the antinodal region \cite{KondoDisentangling2011}. Moreover, the destruction of the octet-model features near the antinode is consistent with the suppression of the spectral weight \cite{KondoCompetition2009, HashimotoParticle-hole2010} and/or band renormalization in the antinodal region by the pseudogap \cite{HashimotoParticle-hole2010}. The termination point may not be directly related to the antiferromagnetic zone boundary, as reported via STS \cite{KohsakaHow2008}.

In summary, we showed that, unlike the ARPES extracted result, the gap function from AC ARPES as well as FT STS is inaccurate due to the compounded effect of intrinsic spectral widths, the pseudogap and matrix elements. Further, our results suggest the more general importance of such effects when investigating $\mathbf{k}$-space electronic structure from $\mathbf{r}$- and $\mathbf{q}$-space information. This may reconcile ARPES and FT-STS results and suggests that the superconductivity is triggered by the superconducting gap opening on the Fermi arc with a simple $d_{x^2-y^2}$ functional form. This contradicts the STS proposal of  superconductivity with the Fermi arc and with no sharp signature of a superconducting transition. It is likely that the observance of QPI above $T_\mathrm{c}$ may mainly be caused by a gap in the antinodal region and the lack of sensitivity to states near the nodal region.

We thank W.-S. Lee, A. Fujimori, P. Hirschfeld D. Scalapino, E. Abrahams and S. Kivelson for helpful discussions and Y. Li for experimental assistance on SQUID measurements. R.-H.H. thanks the SGF for financial support. This work is supported by the Department of Energy, Office of Basic Energy Science under Contract No. DE-AC02-76SF00515.


\begin{thebibliography}{10}

\bibitem{TanakaDistinct2006}
K. Tanaka {\it et~al.}, Science {\bf 314},  1910  (2006).

\bibitem{LeeAbrupt2007}
W.~S. Lee {\it et~al.}, Nature {\bf 450},  81  (2007).

\bibitem{KondoCompetition2009}
T. Kondo {\it et~al.}, Nature {\bf 457},  296  (2009).

\bibitem{KanigelProtected2007}
A. Kanigel {\it et~al.}, Phys. Rev. Lett. {\bf 99},  157001  (2007).

\bibitem{ChatterjeeObservation2010}
U. Chatterjee {\it et~al.}, Nature Phys. {\bf 6},  99  (2009).

\bibitem{DamascelliAngle-resolved2003}
A. Damascelli, Z. Hussain, and Z.-X. Shen, Rev. Mod. Phys. {\bf 75},  473
  (2003).

\bibitem{HoffmanImaging2002}
J.~E. Hoffman {\it et~al.}, Science {\bf 297},  1148  (2002).

\bibitem{McElroyRelating2003}
K. McElroy {\it et~al.}, Nature {\bf 422},  592  (2003).

\bibitem{HanaguriQuasiparticle2007}
T. Hanaguri {\it et~al.}, Nature Phys. {\bf 3},  865  (2007).

\bibitem{WiseImaging2009}
W.~D. Wise {\it et~al.}, Nature Phys. {\bf 5},  213  (2009).

\bibitem{KohsakaHow2008}
Y. Kohsaka {\it et~al.}, Nature {\bf 454},  1072  (2008).

\bibitem{LeeSpectroscopic2009}
J. Lee {\it et~al.}, Science {\bf 325},  1099  (2009).

\bibitem{WangQuasiparticle2003}
Q.~H. Wang and D.~H. Lee, Phys. Rev. B {\bf 67},  020511  (2003).

\bibitem{MarkiewiczBridging2004}
R.~S. Markiewicz, Phys. Rev. B {\bf 69},  214517  (2004).

\bibitem{WulinModel2009}
D. Wulin, Y. He, C.-C. Chien, D.~K. Morr and K. Levin, Phys. Rev. B {\bf 80},  134504  (2009).

\bibitem{WulinContrasting}
D. Wulin, C.-C. Chien, D.~K. Morr, and K. Levin, Phys. Rev. B (R) {\bf 81},  100504   (2010).

\bibitem{AndersenLDA1995}
O.~K. Andersen, A. Liechtenstein, O. Jepsen, and F. Paulsen, J. Phys. Chem.
  Solids {\bf 56},  1573  (1995).

\bibitem{WuAbsence2000}
C. Wu, T. Xiang, and Z.-B. Su, Phys. Rev. B {\bf 62},  14427  (2000).

\bibitem{MartinImpurity2002}
I. Martin, A.~V. Balatsky, and J. Zaanen, Phys. Rev. Lett. {\bf 88},  097003
  (2002).

\bibitem{NieminenOrigin2009}
J. Nieminen, H. Lin, R.~S. Markiewicz, and A. Bansil, Phys. Rev. Lett. {\bf
  102},  037001  (2009).

\bibitem{McElroyElastic2006}
K. McElroy {\it et~al.}, Phys. Rev. Lett. {\bf 96},  067005  (2006).

\bibitem{ChatterjeeNondispersive2006}
U. Chatterjee {\it et~al.}, Phys. Rev. Lett. {\bf 96},  107006  (2006).

\bibitem{SupACARPES}
See supplemental material at http://link.aps.org/supplemental/10.1103/PhysRevLett.106.167003 for additional information.

\bibitem{BansilImportance99}
A. Bansil, and M. Lindroos, Phys. Rev. Lett. {\bf
  83},  5154  (1999). 

\bibitem{Hoffmanfour2002}
J.~E. Hoffman {\it et~al.}, Science {\bf 295},  466  (2002).

\bibitem{WiseCharge-density-wave2008}
W.~D. Wise {\it et~al.}, Nature Phys. {\bf 4},  696  (2008).

\bibitem{KondoDisentangling2011}
T. Kondo {\it et~al.}, Nature Phys. {\bf 7},  21  (2010).

\bibitem{HashimotoParticle-hole2010}
M. Hashimoto {\it et~al.}, Nature Phys. {\bf 6},  414  (2010).

\end{thebibliography}
\end{document}